\documentclass{pasa}

\title[Light curve solutions of eccentric \emph{Kepler} binaries]{Light curve solutions of ten eccentric \emph{Kepler }binaries, three of them with tidally induced humps}
\author[D. Kjurkchieva and D. Vasileva]{D. Kjurkchieva$^1$ \and D. Vasileva$^{1}$\thanks{database of the \emph{Kepler} mission}\\
\affil{$^1$Department of Physics, Shumen University, 9700 Shumen, Bulgaria}%
\affil{$^2$Department of Physics, Shumen University, 9700 Shumen, Bulgaria}}%
\jid{PASA}
\doi{10.1017/pas.\the\year.xxx}
\jyear{\the\year}

\begin{document}%
\begin{abstract}

We carried out light curve solutions of ten detached
eclipsing eccentric binaries observed by \emph{Kepler}.
The formal errors of the derived parameters from the light curve
solutions are below 1 $\%$. Our results give indications that the
components of the eccentric binaries (especially those with mass
ratios below 0.5) do not follow precisely the empirical relations
between the stellar parameters derived from the study of
circular-orbit binaries. We found the following peculiarities of
the targets: (a) the components of KIC 9474969 have almost the
same temperatures while their radii and masses differ by a factor
around 2.5; (b) KIC 6949550 reveals semi-regular light variations
with an amplitude of 0.004 and a period around 7 d which are
modulated by long-term variations; (c) KIC 6220470, KIC 11071207
and KIC 9474969 exhibit tidally induced ''hump'' around the
periastron. These are the targets with the biggest relative radii
of our sample. We derived the dependence of the hump amplitude on
the relative stellar radii, eccentricity and mass ratio of
eccentric binary consisting of MS stars.
\end{abstract}
\maketitle%
\section{INTRODUCTION }
\label{sec:intro}

The most studies on binaries have focused on near-circular orbits
(Pichardo et al. 2005). However, it is generally supposed that
eccentric binaries are mainly produced as a result from the
fragmentation (Bonnell $\&$ Bastien 1992, Bate 1997; Bate $\&$
Bonnell 1997). Moreover, the empirical data reveal that the
main-sequence binary systems typically have eccentric orbits
(Duquennoy $\&$ Mayor 1991). The probable reason the eccentric
binaries to be poorly studied is that they are wide stellar
systems with long periods (Bate et al. 2002) requiring prolonged
observations. Recently this condition was satisfied by the huge
surveys as ROTSE, MACHO, ASAS, SuperWASP, etc. The next important
step was made by the space mission \emph{Kepler} (Koch et al.
2010). Due to its extended and nearly uninterrupted data set above
thousand detached systems were discovered, considerable part of
them on eccentric orbits.

\emph{Kepler} did not only provide many new eccentric binaries
but the unique precision of its observational data transformed
these binary stars from objects of the celestial mechanics to an
important field of the stellar astrophysics. They became probes
for study of the tidal phenomena: mechanisms for circularization
of the orbits and synchronization of the stellar rotation with the
orbital motion; impermanent mass transfer occurring close to the
periastron (Sepinsky et al. 2007a, Lajoie $\&$ Sills 2011);
apsidal motion; tidally excited brightening and oscillations.

The theoretical studies reveal that the secular changes of the
orbital separation and eccentricity could be positive or negative
(depending on the mass ratio and eccentricity), and could occur on
timescales ranging from a few million years to a few billion years
(Sepinsky et al. 2007b, 2009). Thus, the widespread assumption for
rapid circularization becomes inapplicable, i.e. binaries can
remain on eccentric orbits for long periods of time. Hence, the
binary stars on eccentric orbits have important evolutional role.

The eclipsing eccentric binaries (EEBs) with an apsidal motion
provide an important observational test of the theoretical models
of stellar structure and evolution (Kopal 1978, Claret $\&$
Gimenez 1993). The coefficients of internal structure $k_j$ are
used to describe the external potential of a distorted
configuration as a function of its internal structure in the form
of series expansion (Claret $\&$ Gimenez 1991). In most cases only
$k_2$ is important since higher harmonics in the apsidal motion
decrease rapidly. In order to compare with observations an average
value $\bar{k}_2=(c_{1}k_{12}+c_{2}k_{22})/(c_{1}+c_{2})$ of the
contribution of star 1 and 2 ($k_{12}$ and $k_{22}$) is used. The
coefficients $c_{1}$ and $c_{2}$ are known functions of observable
stellar and orbital parameters (eccentricity, radii, masses and
rotational velocities). The individual contributions can be
separated in two terms, rotational and tidal. In most cases tidal
terms dominate the apsidal motions of binary systems (Claret $\&$
Gimenez 1993). Hence, the ''apsidal motion test'' could be used as
a probe of stellar internal structure if precise values of global
parameters are available (Claret $\&$ Gimenez 2010). There are
important recent studies of this effect based on the observed
apsidal motion of double-lined binaries (Lacy et al. 2015, Zasche
et al. 2014, Bulut et al. 2014, Garcia et al. 2014, Kozyreva $\&$
Kusakin 2014, Harmanec et al. 2014, Wilson $\&$ Van Hamme 2014,
Hambleton et al. 2013, Wolf et al. 2013, Claret 2012, Zasche 2012,
Kuznetsov et al. 2011, Claret $\&$ Gimenez 2010, Wolf et al. 2010,
Gimenez $\&$ Quintana 1992, Barembaum $\&$ Etzel 1995, etc.).

In addition to the classical Newtonian contribution, the observed
apsidal motion includes term of the General Relativity
(Levi-Civita 1937, Gimenez 1985). Information for the
apsidal motions of 128 targets in our Galaxy can be found in the
catalog of Petrova $\&$ Orlov (1999). The study of EEBs in close
galaxies SMC and LMC began in the new millenium (Graczyk 2003;
Michalska $\&$ Pigulski 2005; Michalska 2007; Bulut $\&$ Demircan
2007; North et al. 2010; Zasche $\&$ Wolf 2013; Zasche et al.
2014).

Probably the most amazing peculiarities of the binary stars on
eccentric orbits are the tidally excited oscillations (harmonics
of the orbital period) and brightening around the periastron. They
were theoretically predicted by Kumar et al. (1995) to explain the
$\sim$1 day oscillations with amplitude of 0.002 of B-type star
orbiting a neutron star (radio pulsar PSR 0045-7319 with
\emph{e}=0.81 and P$_{orb}$=51 d). Further this analytic theory
was used for explanation of: (a) the oscillation with frequency of
exactly ten times the orbital frequency of the slowly pulsating
B-star in the binary HD177863 (De Cat et al. 2000); (b) the
oscillations of the A-type primary of the eccentric binary HD
209295 (Handler et al. 2002); (c) the oscillations of the
eccentric, short-period early-type binary HD 174884 (Maceroni et
al. 2009).

The theory of the tidally excited phenomena further was developed
by Willems (2003), Zahn (2005), Willems $\&$ Claret (2005),
Willems (2007), Hernandez-Gomez et al. (2011), Gundlach $\&$
Murphy (2011), Burkart et al. (2012), Song et al. (2013),
Borkovits et al. (2014), etc.

Brilliant confirmations of the theoretical predictions of Kumar et
al. (1995) were discovered by the \emph{Kepler} mission. KOI 54
exhibits two remarkable features: a periodic brightening spike of
~0.7 $\%$ occurring at the periastron and a ~0.1 $\%$ ''beat''
pattern of pulsations in phase with the brightening events (Welsh
et al. 2011, Burkart et al. 2012). Thompson et al. (2012)
discovered the next 16 similar objects (most of them noneclipsing)
in the \emph{Kepler} archive and called them ''heartbeat'' stars
due to their shape of light variability reminiscent an
echocardiogram. Their light curves are different: some dim before
they brighten, others dim after they brighten, and others show
distinct 'W' or 'M' shapes. Stellar oscillations at harmonics of
the heartbeat periods of some of these targets were also found.

Answering to the appeal to use the available resources of the
{\emph{Kepler} database for additional research we undertook study
of some types of binary systems from the EB catalog (Dimitrov et
al. 2012, Kjurkchieva $\&$ Dimitrov 2015). The goal of this study
was to obtain the orbits and parameters of ten eccentric binaries
based on the \emph{Kepler} data as well as to search for tidally
induced phenomena.

\begin{table*}
\begin{minipage}[t]{\textwidth}
\caption{Parameters of the targets from the EB catalog
(period\emph{ P}, \emph{Kepler} magnitude \emph{kmag}, mean
temperature $T_m$, widths $w_i$ and depths $d_i$ of the eclipses)
and phases of the secondary eclipses $\varphi_2$} \label{tab:log1}
\centering
\begin{scriptsize}
\renewcommand{\footnoterule}{}
\begin{tabular}{ccccccccc}
\hline\hline
Star            &   \emph{P} [d]& \emph{kmag }& $T_m$    &  $w_1$   &  $w_2$   &   $d_1$   &   $d_2$  &  $\varphi_2$   \\
\hline
KIC 6220470     &   8.144136    &   13.960  &    7377    &  0.045   &   0.037   &   0.091   &   0.010  &   0.400       \\
KIC 8296467     &   10.33356    &   15.177  &    5316    &  0.024   &   0.016   &   0.531   &   0.326  &   0.625      \\
KIC 6877673     &   36.758887   &   13.676  &    5819    &  0.016   &   0.013   &   0.273   &   0.188  &   0.439      \\
KIC 9658118     &   24.059995   &   14.158  &    6225    &  0.017   &   0.035   &   0.473   &   0.457  &   0.603       \\
KIC 12306808    &   37.878484   &   13.265  &    5738    &  0.011   &   0.013   &   0.220   &   0.169  &   0.555       \\
KIC 5553624     &   25.762071   &   14.231  &    5358    &  0.023   &   0.007   &   0.449   &   0.246  &   0.420      \\
KIC 9474969     &   21.570506   &   12.462  &    6085    &  0.027   &   0.023   &   0.133   &   0.116  &   0.759       \\
KIC 11391181    &   8.617340    &   15.257  &    5218    &  0.017   &   0.019   &   0.181   &   0.106  &   0.607       \\
KIC 11071207    &   8.049635    &   13.831  &    6427    &  0.031   &   0.040   &   0.279   &   0.140  &   0.621       \\
KIC 6949550     &   7.841067    &   15.144  &    5719    &  0.033   &   0.031   &   0.355   &   0.350  &   0.332      \\
\hline\hline
\end{tabular}
\end{scriptsize}
\end{minipage}
\end{table*}

\section{SELECTION OF THE TARGETS}

Above two thousands eclipsing binaries (EBs) have been identified
and included in the \emph{Kepler} EB catalog (Prsa et al. 2011;
Slawson et al. 2011), around 1261 of them have been initially
classified as detached systems. An automate fitting of the light
curves have been used for determination of their ephemerides.

We reviewed visually the \emph{Kepler} EB catalog to search for
detached binaries with eccentric orbits and found around 250
targets which phase difference between the primary and secondary
minimum differed considerably from 0.5 and which durations of the
two light minima were not equal. Most of the found light curves
were with quite narrow eclipses (due to the long periods). We
chose to model ten binaries with relatively long eclipses (above
0.01 in phase units) allowing precise light curve solutions.

Table 1 presents available information for the targets (Prsa et
al. 2011): orbital period $P$; \emph{Kepler} magnitude
\emph{kmag}; mean temperature $T_m$; width of the primary eclipse
w$_1$ (in phase units); width of the secondary eclipse w$_2$ (in
phase units); depth of the primary eclipse d$_1$ (in flux units);
depth of the secondary eclipse d$_2$ (in flux units). We added to
the available target data in Table 1 the phases $\varphi_2$ of
their secondary eclipses (the phases $\varphi_1$ of the primary
eclipses are 0.0).

The detailed review of the \emph{Kepler} data of our targets
revealed that their light curves did not change during the
different cycles and observational quarters. This allows to model
any data set. We chosen to use several consecutive cycles from the
middle quarters of each target.

\begin{figure}
   \centering
   \includegraphics[width=0.99\columnwidth]{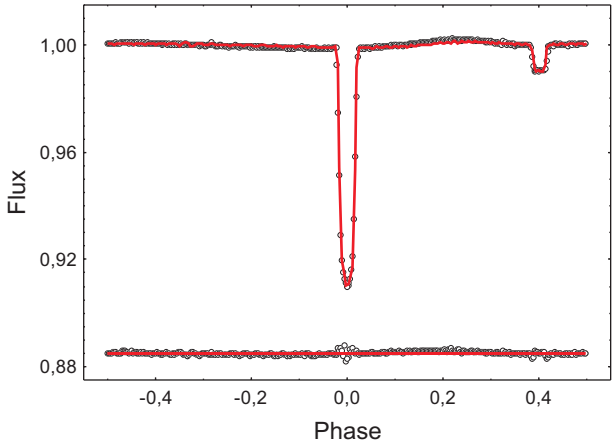}
  % \begin{minipage}[]{85mm}
   \caption{Top: the folded light curve of KIC 6220470 and its fit; Bottom: the corresponding residuals
   (shifted vertically by different number to save space).
   Color version of this figure is available in the online journal.}
%\end{minipage}
   \label{Fig1}
   \end{figure}

\begin{figure}
   \centering
   \includegraphics[width=8cm, angle=0]{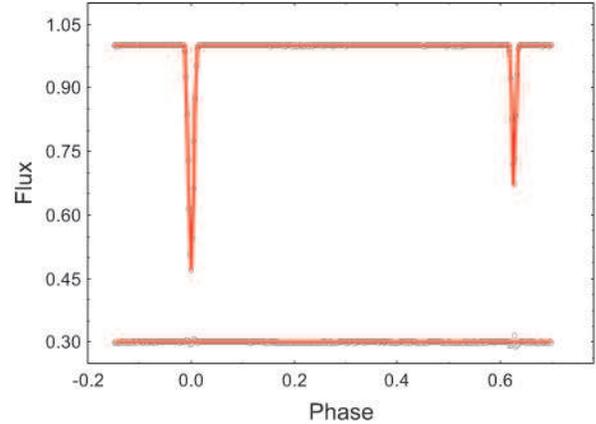}
  % \begin{minipage}[]{85mm}
   \caption{Same as Fig. 1 for KIC 8296467}
%\end{minipage}
   \label{Fig2}
   \end{figure}

\begin{figure}
   \centering
   \includegraphics[width=8cm, angle=0]{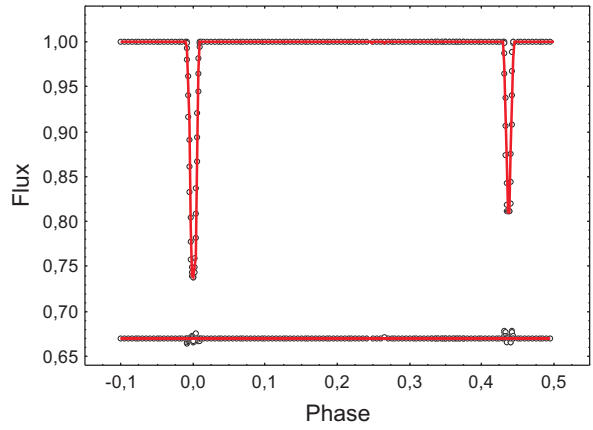}
  % \begin{minipage}[]{85mm}
   \caption{Same as Fig. 1 for KIC 6877673}
%\end{minipage}
   \label{Fig3}
   \end{figure}

\begin{figure}
   \centering
   \includegraphics[width=8cm, angle=0]{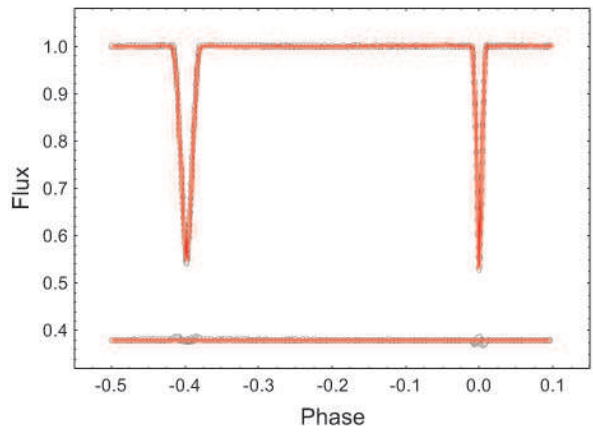}
  % \begin{minipage}[]{85mm}
   \caption{Same as Fig. 1 for KIC 9658118}
%\end{minipage}
   \label{Fig4}
   \end{figure}

\begin{figure}
   \centering
   \includegraphics[width=8cm, angle=0]{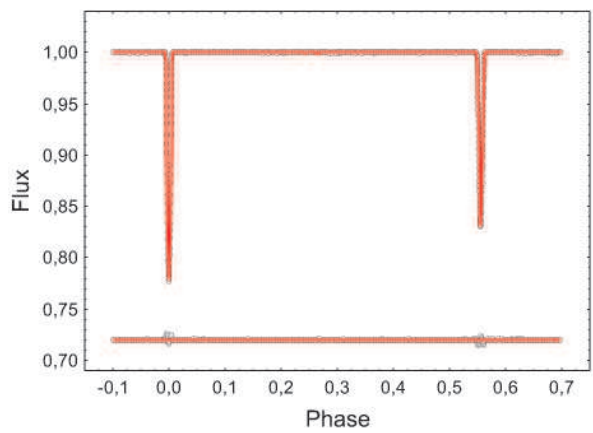}
  % \begin{minipage}[]{85mm}
   \caption{Same as Fig. 1 for  KIC 12306808}
%\end{minipage}
   \label{Fig5}
   \end{figure}

\begin{figure}
   \centering
   \includegraphics[width=8cm, angle=0]{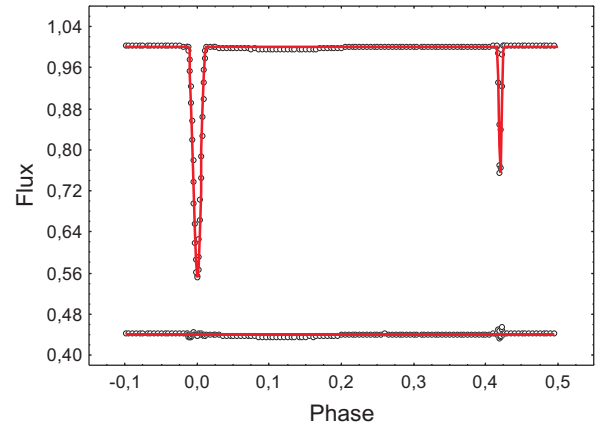}
  % \begin{minipage}[]{85mm}
   \caption{Same as Fig. 1 for KIC 5553624}
%\end{minipage}
   \label{Fig6}
   \end{figure}

\begin{figure}
   \centering
   \includegraphics[width=8cm, angle=0]{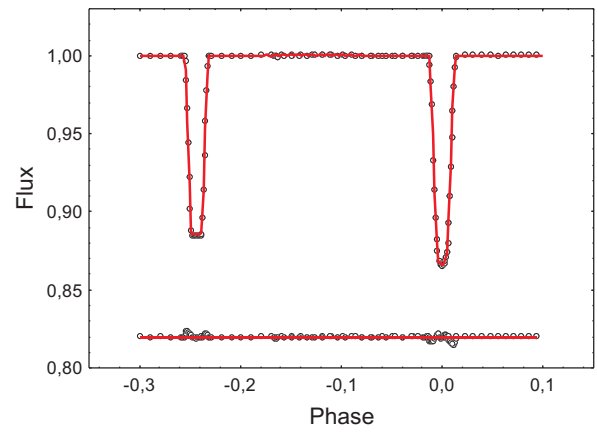}
  % \begin{minipage}[]{85mm}
   \caption{Same as Fig. 1 for KIC 9474969}
%\end{minipage}
   \label{Fig7}
   \end{figure}

\begin{figure}
   \centering
   \includegraphics[width=8cm, angle=0]{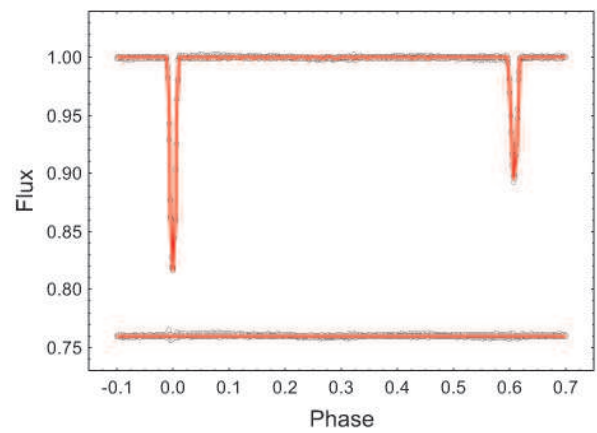}
  % \begin{minipage}[]{85mm}
   \caption{Same as Fig. 1 for KIC 11391181}
%\end{minipage}
   \label{Fig8}
   \end{figure}

\begin{figure}
   \centering
   \includegraphics[width=8cm, angle=0]{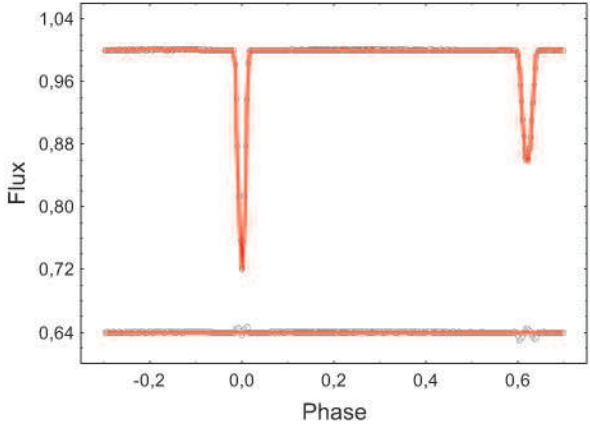}
  % \begin{minipage}[]{85mm}
   \caption{Same as Fig. 1 for KIC 11071207}
%\end{minipage}
   \label{Fig9}
   \end{figure}

\begin{figure}
   \centering
   \includegraphics[width=8cm, angle=0]{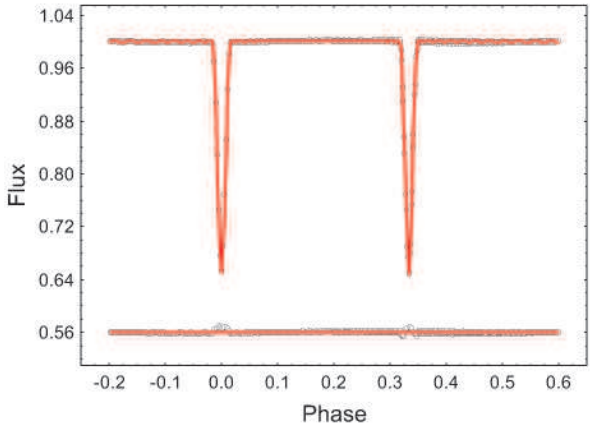}
  % \begin{minipage}[]{85mm}
   \caption{Same as Fig. 1 for KIC 6949550}
%\end{minipage}
   \label{Fig10}
   \end{figure}

\begin{table}
\begin{minipage}[t]{\columnwidth}
\caption{The derived orbital parameters of the targets:
eccentricity \emph{e}, periastron angle $\omega$ and periastron
phase $\varphi_{per}$} \label{tab:log1} \centering
\begin{scriptsize}
\renewcommand{\footnoterule}{}
\begin{tabular}{cccc}
\hline\hline
Star            &  \emph{e }        &   $\omega$ [deg]  & $\varphi_{per}$  \\
\hline
KIC 6220470     &   0.1879 $\pm$ 0.0002  &   215.03 $\pm$ 0.01 & 0.294     \\
KIC 8296467     &   0.2781 $\pm$ 0.0001  &   314.48 $\pm$ 0.01 & 0.696     \\
KIC 6877673     &   0.1704 $\pm$ 0.0002  &   235.26 $\pm$ 0.01 & 0.369     \\
KIC 9658118     &   0.3723 $\pm$ 0.0001  &   66.01  $\pm$ 0.01 & 0.971     \\
KIC 12306808    &   0.1089 $\pm$ 0.0001  &   36.74  $\pm$ 0.01 & 0.878     \\
KIC 5553624     &   0.5204 $\pm$ 0.0001  &   258.12 $\pm$ 0.01 & 0.411     \\
KIC 9474969     &   0.4165 $\pm$ 0.0001  &   354.78 $\pm$ 0.01 & 0.867     \\
KIC 11391181    &   0.1797 $\pm$ 0.0001  &   18.94  $\pm$ 0.01 & 0.854     \\
KIC 11071207    &   0.2256 $\pm$ 0.0004  &   33.00 $\pm$ 0.01  & 0.897     \\
KIC 6949550     &   0.2661 $\pm$ 0.0001  &   186.64 $\pm$ 0.01 & 0.183     \\
\hline\hline
\end{tabular}
\end{scriptsize}
\end{minipage}
\end{table}

\begin{table*}
\begin{minipage}[t]{\textwidth}
\caption{Parameters of the best light curve solutions: orbital
inclination\emph{ i}, mass ratio \emph{q}, temperatures $T_i$,
relative radii $r_i$ and relative luminosities $l_i$ of the
stellar components}
 \label{tab:log1}
\centering
\begin{scriptsize}
\renewcommand{\footnoterule}{}
\begin{tabular}{ccccccccc}
\hline\hline
Star            &   \emph{i} &   \emph{q  }  &   $T_1$ [K] &  $T_2$ [K] &   $r_1$          &   $r_2$       & $l_{1}$ & $l_{2}/l_{1}$   \\
\hline
KIC 6220470     &   89.50    &   0.446       &   7875      &   4382      &   0.1014         &   0.0286      &  0.9918  &  0.0082  \\
                &  $\pm$0.01 &   $\pm$0.001  &   $\pm$267  &   $\pm$70   &   $\pm$0.0010    &   $\pm$0.0005 &          &          \\

KIC 8296467     &   89.58    &   0.776       &   5726      &   5077      &   0.0344         &   0.0325      &  0.6478  &  0.5436  \\
                &  $\pm$0.01 &   $\pm$0.001  &   $\pm$8    &   $\pm$6    &   $\pm$0.0003    &   $\pm$0.0004 &          &          \\

KIC 6877673     &   89.71    &   0.798       &   6117      &   5706      &   0.0312         &   0.0169      &  0.8234  &  0.2144  \\
                &  $\pm$0.01 &   $\pm$0.004  &   $\pm$29   &   $\pm$25   &   $\pm$0.0010    &   $\pm$0.0010 &          &          \\

KIC 9658118     &   89.93    &   0.973       &   6212      &   6193      &   0.0431         &   0.0396      &  0.5445  &  0.8365  \\
                &  $\pm$0.01 &   $\pm$0.002  &   $\pm$3    &   $\pm$3    &   $\pm$0.0002    &   $\pm$0.0001 &          &          \\

KIC 12306808    &   88.668   &   0.807       &   5772      &   5693      &   0.0249         &   0.0214      &  0.5910  &  0.6920  \\
                &  $\pm$0.001&   $\pm$0.001  &   $\pm$4    &   $\pm$3    &   $\pm$0.0001    &   $\pm$0.0004 &          &          \\

KIC 5553624     &   89.623   &   0.682       &   6122      &   5120      &   0.0244         &   0.0200      &  0.7772  &  0.2866  \\
                &  $\pm$0.002&   $\pm$0.001  &   $\pm$8    &   $\pm$5    &   $\pm$0.0003    &   $\pm$0.0005 &          &          \\

KIC 9474969     &   87.94    &   0.403       &   5737      &   5574      &   0.0679         &   0.0258      &  0.8876  &  0.1266  \\
                &   $\pm$0.01&   $\pm$0.002  &   $\pm$47   &   $\pm$45   &   $\pm$0.0001    &   $\pm$0.0050 &          &          \\

KIC 11391181    &   87.42    &   0.7728      &   5199      &   4778      &   0.04805        &   0.03125     &  0.7868  &  0.2709  \\
                &  $\pm$0.01 &   $\pm$0.0014 &   $\pm$5    &   $\pm$4    &   $\pm$0.00030   &   $\pm$0.00010&          &          \\

KIC 11071207    &   87.47    &   0.4304      &   6836      &  5913       &   0.0793         &   0.0446      &  0.8518  &  0.2861  \\
                &   $\pm$0.01&   $\pm$0.0020 &   $\pm$24   &  $\pm$19    &   $\pm$0.0007    &   $\pm$0.0013 &          &          \\

KIC 6949550     &   88.45    &   0.9859      &   5802      &   5784      &   0.0504         &   0.0502      &  0.5058  &  0.9770  \\
                &  $\pm$0.01 &   $\pm$0.0018 &   $\pm$6    &   $\pm$5    &   $\pm$0.0008    &   $\pm$0.0007 &          &          \\
\hline\hline
\end{tabular}
\end{scriptsize}
\end{minipage}
\end{table*}

\section{LIGHT CURVE SOLUTIONS}

We carried out the modeling of the \emph{Kepler} data by the
binary modelling package \emph{PHOEBE} (Prsa $\&$ Zwitter 2005).
It is based on the Wilson--Devinney (WD) code (Wilson $\&$
Devinney 1971, Wilson 1979, Wilson $\&$ Van Hamme 2004).
\emph{PHOEBE} incorporates all the functionality of the WD code
but also provides a graphical user interface alongside other
improvements, including updated filters for the various recent
space missions as \emph{Kepler}. That is why \emph{PHOEBE} is
highly appropriate for modeling of the precise Kepler data
(Hambleton et al. 2013).

We used several considerations for the light curve solutions.

(1) The almost flat out-of-eclipse parts of the observed light
curve gave us the grounds to use the detached mode of modeling.

(2) The light curve solution of eccentric binaries by
\emph{PHOEBE} also by each light curve synthesis software) without
any guessed values of the eccentricity $e$ and periastron angle
$\omega$ is quite time-consuming task. That is why we calculated
preliminary values of these orbit parameters by the formulae
\begin{equation}
e_0 \cos\omega_0=\frac{\pi}{2}[(\varphi_2-\varphi_1)-0.5]
\end{equation}
\begin{equation}
e_0 \sin\omega_0=\frac{w_2-w_1}{w_2+w_1}
\end{equation}
They were obtained as approximations of formulae (9-25) and (9-37)
of Kopal (1978). We calculated $e_0$ and $\omega_0$ by our
expressions (1-2) using the values of $\varphi_2$, $w_1$ and $w_2$
from Table 1 ($\varphi_1$=0). The obtained values of $e_0$ and
$\omega_0$ were used as input parameters of \emph{PHOEBE}.

\begin{figure}
   \centering
   \includegraphics[width=5.5cm, angle=0]{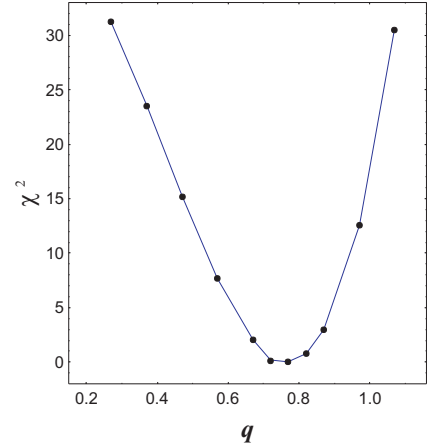}
  % \begin{minipage}[]{85mm}
   \caption{Sensibility of our light curve solution of KIC 11391181 (measured by $\chi^2$) to the mass ratio
   (the rest parameters last fixed at their final values)}
%\end{minipage}
   \label{Fig11}
   \end{figure}

(3) The mean temperatures $T_m$ of our targets (Table 1) allowed
us to adopt coefficients of gravity brightening 0.32 and
reflection effect 0.5 appropriate for stars with convective
envelopes. The only exception was the hot primary of KIC 6220470
requiring parameters of star with radiative envelope (gravity
brightening 1.0 and reflection effect 1.0). The contribution of
the gravity brightening effect (and its coefficients) is not
considerable for detached binaries as our targets.

(4) Initially the synchronicity parameters were kept fixed at
values of unity.

(5) Initially the primary temperature $T_{1}$ was fixed to be equal to
the mean target temperature $T_{m}$ (Table 1) that has been
estimated using dedicated pre-launch ground-based optical
multi-color photometry plus Two Micron All Sky Survey (2MASS) J,
K, and H magnitudes (Prsa et al. 2011).

%The coefficients were interpolated from the van Hamme's tables (van Hamme 1993).
%In order to be adequate to the \emph{Kepler} filter we used the last version of the code: \emph{PHOEBE 3.1}.

Our procedure of the light curve solutions was quite similar to
that of Hambleton et al. (2013) and was carried out in several
stages.

At the very beginning we input some guessed values of the
secondary temperature $T_{2}$, mass ratio $q$, orbital inclination
$i$ and potentials $\Omega_{1,2}$ and varied only the eccentricity
\emph{e} and periastron angle $\omega$ around their input values
$e_0$ and $\omega_0$ to search for the best fit of the phases of
the eclipses estimated by the value of $\chi^2$}. It turned out
that the final values \emph{e} and $\omega$ differed from the
input values up to 10 $\%$. This means that the approximated
formulae (1-2) could be used successfully for calculation of the
input values of eccentricity and periastron angle of eccentric
binaries.

At the second stage we fixed \emph{e} and $\omega$ and varied
simultaneously $T_{2}$, $q$, $i$ and $\Omega_{1,2}$ (and thus
relative radii $r_{1,2}$). We used linear limb-darkening law with
limb-darkening coefficients corresponding to the stellar
temperatures and \emph{Kepler} photometric system (Claret $\&$
Bloemen 2011).

Finally, to adjust the stellar temperatures $T_{1}$ and $T_{2}$
around the mean value $T_m$ we used the procedure described in
Dimitrov $\&$ Kjurkchieva (2015).

The final parameters of the eccentric orbits are given in Table 2
while Table 3 contains the parameters of the stellar
configurations. The synthetic curves corresponding to the
parameters of our light curve solutions are shown in Figs. 1-10 as
continuous lines.

Any surface spots and third lights were not necessary to reproduce
the photometric data and thus the condition the synchronicity
parameters to be fixed at unity turned out without consequence.

The errors in Tables 2-3 are the formal \emph{PHOEBE} errors. Most
of them are smaller than 1 $\%$, excluding the temperatures of the
stellar components of KIC 6220470 whose errors exceed 3 $\%$. We
attributed these bigger errors to the considerably shallower
eclipses of this target, especially the secondary one (see Table
1). The small errors of the derived parameters by our light curve
solutions are natural consequence of the high precision of the
\emph{Kepler} data.

However, one can see two imperfections of the reproducing of the data.

The first one are the relative bigger residuals during the
eclipses (Figs. 1-10). Similar behavior could be seen also for
other \emph{Kepler} binaries (Hambleton et al. 2013, Lehmann et
al. 2013, Maceroni et al. 2014). We attributed them to the effects
of finite integration time studied by Kipping (2010): (i) The
large integration times smear out the light curve signal into a
broader shape (see figure 1 of Kipping, 2010): the detected
ingress  and egress durations are bigger than their natural values
(introducing an additional curvature into the eclipse wings) and
the apparent positions of the contact points are temporally
shifted from their true value; (ii) The large integration times
smear out the curvature of the eclipse bottom caused by the
limb-darkening effect. The two effects are inherent to the
long-cadence \emph{Kepler} data of our targets used for modeling.
Coughlin et al. (2011) also found that the long-cadence
(integration time of 29.43 minutes) \emph{Kepler} data of systems
with small sum of relative radii significantly alter the
morphological shape of a light curve.

The second imperfection are the small ripples on the flat
out-of-eclipse sections of some synthetic light curves as well as
those on the flat bottom of the total eclipses. We attributed them
to numerical shortcoming of the software for the cases of binaries
with small relative radii of the components. Such effect (with
amplitude below 0.1 mmag) was firstly noted by Maceroni et al.
(2009) and interpreted by the description of stellar surfaces with
a finite number of elements. The result of the surface
discretisation seems bigger (with amplitude around 0.2 mmag) for
our targets which components have small relative radii. In fact,
this is the reason the traditional software for light curve
synthesis to be not applicable for modeling of exoplanet transits
(see for instance \emph{PHOEBE} \emph{scientific} \emph{reference}
by A. Prsa, 2011).

We estimated the global parameters of the target components (Table
4) using the empirical statistical relations mass-temperature,
radius-temperature and luminosity-temperature for MS stars
(Boyajian et al. 2013) which were approximated by combinations of
linear functions. Thus, the absolute parameters in Table 4 are result of
interpolation of empirical relations, not derived from
observational data, and the listed errors are due to the
interpolation process.

\begin{table*}
\begin{minipage}[t]{\textwidth}
\caption{Masses $M_i$, radii $R_i$ and luminosities $L_i$ of the
target components (in solar units) according to the empirical
relations. \textbf{Their errors are due to the interpolation
process. }} \label{tab:log1} \centering
\begin{scriptsize}
\renewcommand{\footnoterule}{}
\begin{tabular}{ccccccc}
\hline\hline
Star            &     $M_1$        &    $M_2 $      &     $R_1$     &   $R_2$       &    $L_1$        &   $L_2$   \\
\hline
KIC 6220470     &   1.75$\pm$0.09  &  0.80$\pm$0.01 & 1.95$\pm$0.08 & 0.73$\pm$0.09 &  10.9$\pm$2.2   &  0.18$\pm$0.02   \\
KIC 8296467     &   1.00$\pm$0.01  &  0.88$\pm$0.01 & 1.20$\pm$0.01 & 1.04$\pm$0.01 &  1.47$\pm$0.01  &  0.52$\pm$0.01  \\
KIC 6877673     &   1.05$\pm$0.01  &  0.95$\pm$0.01 & 1.31$\pm$0.01 & 1.19$\pm$0.01 &  2.63$\pm$0.09  &  1.41$\pm$0.04 \\
KIC 9658118     &   1.09$\pm$0.01  &  1.08$\pm$0.01 & 1.33$\pm$0.01 & 1.32$\pm$0.01 &  3.01$\pm$0.01  &  3.01$\pm$0.01 \\
KIC 12306808    &   0.96$\pm$0.01  &  0.95$\pm$0.01 & 1.20$\pm$0.01 & 1.18$\pm$0.01 &  1.51$\pm$0.01  &  1.38$\pm$0.01 \\
KIC 5553624     &   1.05$\pm$0.01  &  0.86$\pm$0.01 & 1.31$\pm$0.01 & 1.05$\pm$0.01 &  2.63$\pm$0.02  &  0.56$\pm$0.01   \\
KIC 9474969     &   1.06$\pm$0.02  &  0.94$\pm$0.01 & 1.32$\pm$0.01 & 1.16$\pm$0.02 &  1.47$\pm$0.09  &  1.14$\pm$0.06  \\
KIC 11391181    &   0.89$\pm$0.01  &  0.85$\pm$0.01 & 1.07$\pm$0.01 & 0.92$\pm$0.01 &  0.61$\pm$0.01  &  0.32$\pm$0.01  \\
KIC 11071207    &   1.40$\pm$0.01  &  0.98$\pm$0.01 & 1.53$\pm$0.01 & 1.24$\pm$0.01 &  5.62$\pm$0.11  &  1.90$\pm$0.04  \\
KIC 6949550     &   0.97$\pm$0.01  &  0.96$\pm$0.01 & 1.21$\pm$0.01 & 1.21$\pm$0.01 &  1.58$\pm$0.01  &  1.51$\pm$0.01\\
\hline\hline
\end{tabular}
\end{scriptsize}
\end{minipage}
\end{table*}

\section{DISCUSSION}

The analysis of the light curve solutions of the chosen ten
eccentric binaries led us to several important results.

(1) We established high sensibility of the solutions (measured by
$\chi^2$) to the mass ratio (Fig. 11). This result differed from
those of Michalska $\&$ Pigulski (2004) and Terrell $\&$ Wilson
(2005) who found that for detached eclipsing binaries equally good
fits can be obtained in a very large range of mass ratios. We
attributed the sensibility of our solutions to the mass ratio to
the following reasons: high precision of the \emph{Kepler} data;
very small relative radii of the stars of our sample; eccentric
orbits of our targets.

(2) The temperatures of the stellar components correspond to
spectral type from late A to middle K with dominance of G type.
This result is expected taking into account that \emph{Kepler}
observed mainly solar type stars.

(3) Our targets have quite big eccentricities ($0.11 \leq e \leq
0.52$). This result is due to the selection effect and thus does
not contradict to the expected distribution of \emph{e} peaked at
0.0--0.1 (Prsa et al. 2008).

Our targets do not support the expected relation the eccentricity
to be smaller for the shorter period systems (the shorter period
orbits undergo stronger tidal forces and they circularize more
quickly, leaving a dearth of highly eccentric short period
binaries). Such a trend of increasing of \emph{e} with the period
has been established for evolved MS stars in open cluster M67
(Mathieu $\&$ Mazeh 1988).

(4) The orbital inclinations of the targets are quite near to
90$^0$ (Table 3) that is expected for eclipsing systems with
periods above 8 days. But only two targets, KIC 6220470 and KIC
9474969, undergo total eclipses.

(5) We did not find evidences for apsidal motion of our targets.
The possible reason is the relative short duration of the
\emph{Kepler} observations. Typically apsidal periods are at least
decade long (Michalska $\&$ Pigulski 2005). Moreover, the systems
with apsidal motions are with the shortest orbital periods or with
the largest sum of relative radii for a given eccentricity
(Michalska 2007) but these conditions are not fulfilled for our
targets.

\begin{figure*}
   \centering
   \includegraphics[width=12cm, angle=0]{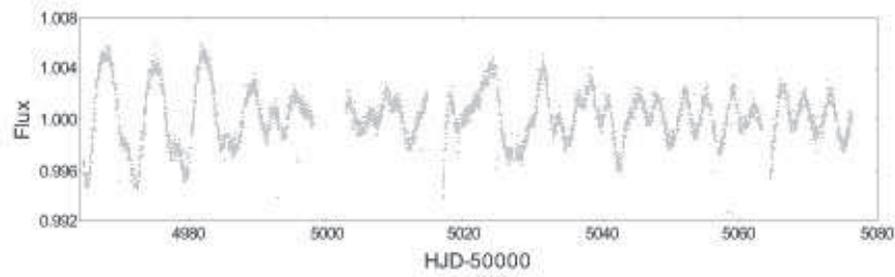}
  % \begin{minipage}[]{85mm}
   \caption{Short-term semi-regular light variations of KIC 6949550}
%\end{minipage}
   \label{Fig12}
   \end{figure*}

    \begin{figure*}
   \centering
   \includegraphics[width=13cm, angle=0]{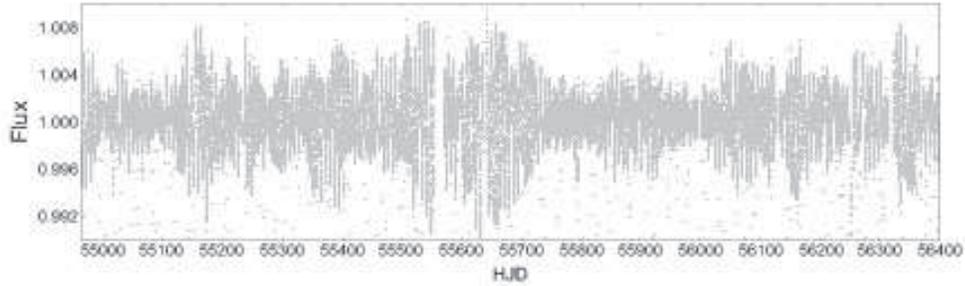}
  % \begin{minipage}[]{85mm}
   \caption{Long-term modulation of the light variations of KIC 6949550}
%\end{minipage}
   \label{Fig13}
   \end{figure*}

(6) The out-of-eclipse light of the targets is constant within 0.2
$\%$. Only KIC 6949550 consisting of two solar type stars reveals
semi-regular light variations (Figs. 12-13) with amplitudes of
0.002--0.004 on timescales of order of 7 d (Fig. 12) which are
modulated with a period around 400 days (Fig. 13). The possibility
these low-amplitude variations to be detected is due to the
unprecedent precision of the \emph{Kepler} data. Their frequency
analysis is object of future study.

(7) The review of the light curves of our targets from different quarters
did not exhibit any long-term variability (excluding KIC 6949550).

(8) The mass ratios of the targets are within the range 0.4--1.0.
North et al. (2010) obtain that \emph{q} for detached systems is
within 0.8--1.1 and for semidetached and contact binaries within
0.4--0.7. Lucy (2006) found an excess of binary systems with $q
\geq$ 0.95 in our Galaxy. Our result implies that probably the
distribution of the mass ratio for eccentric binaries differs from
that for circular-orbit systems.

(9) We established some linear trend between the ratio $r_2/r_1$
and mass ratio \emph{q} (Fig. 14, left) for the ten members of our
small sample.

\begin{figure*}
\begin{center}
\includegraphics[width=5cm]{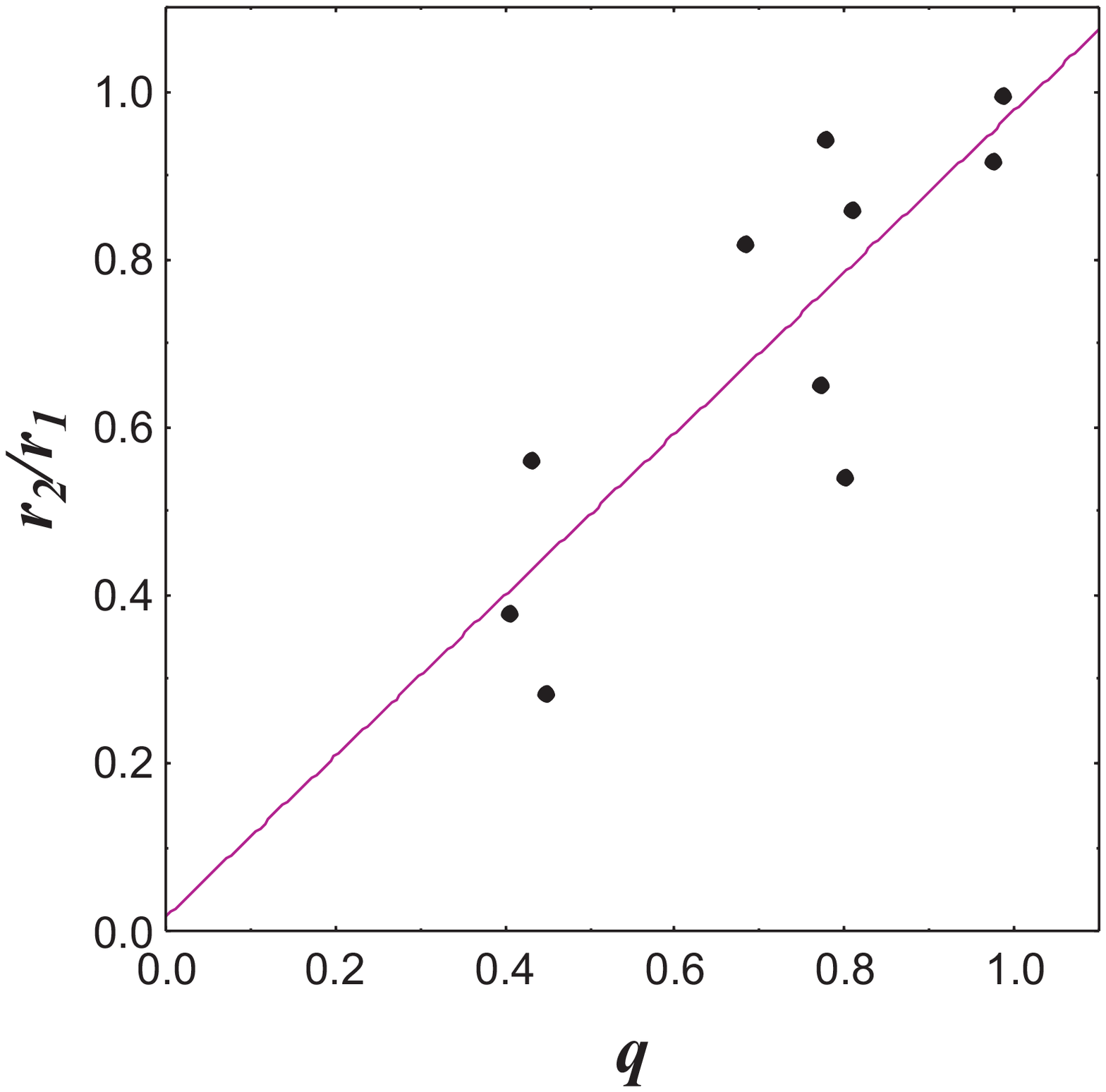}
\includegraphics[width=5cm]{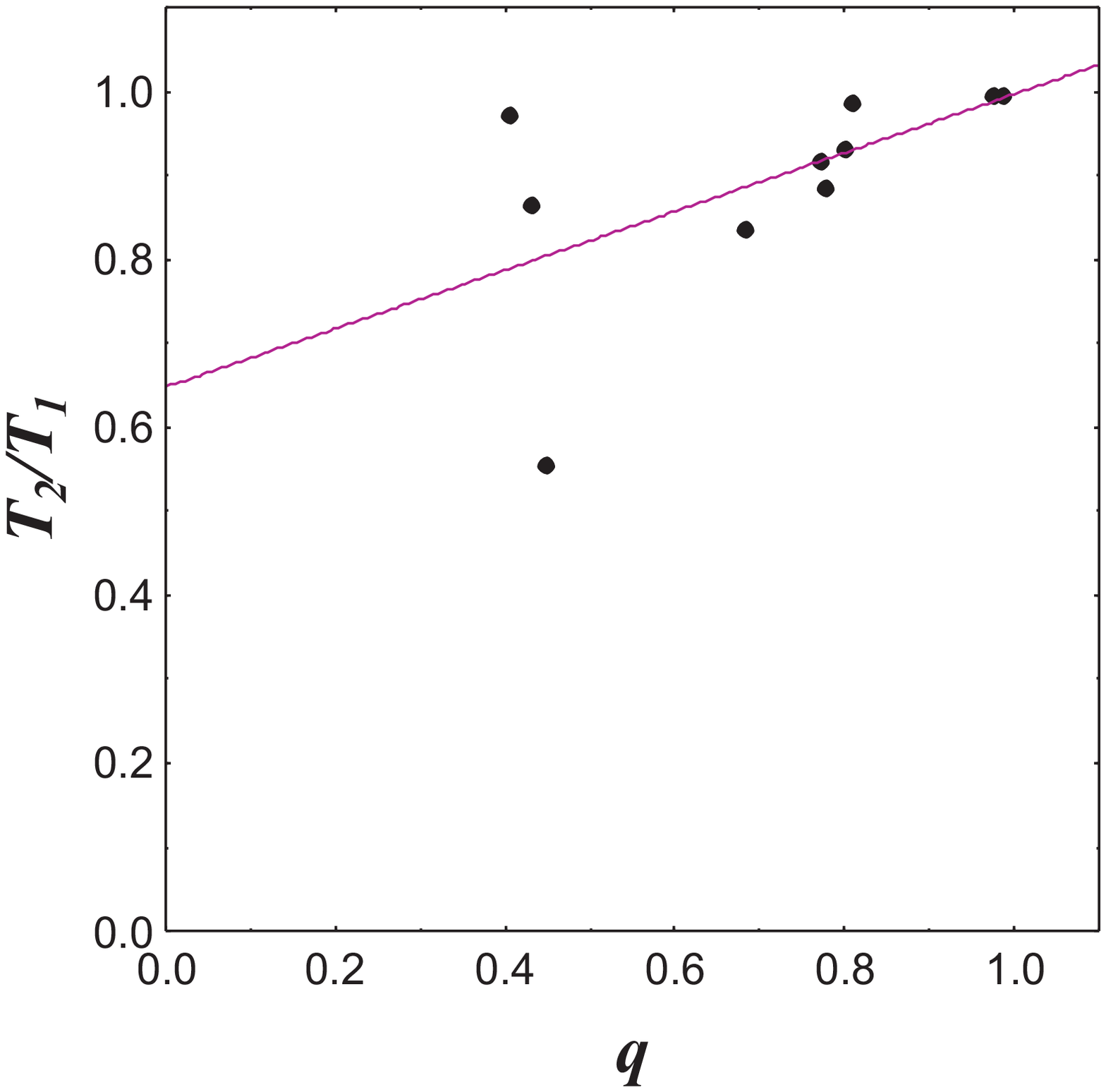}
\includegraphics[width=5cm]{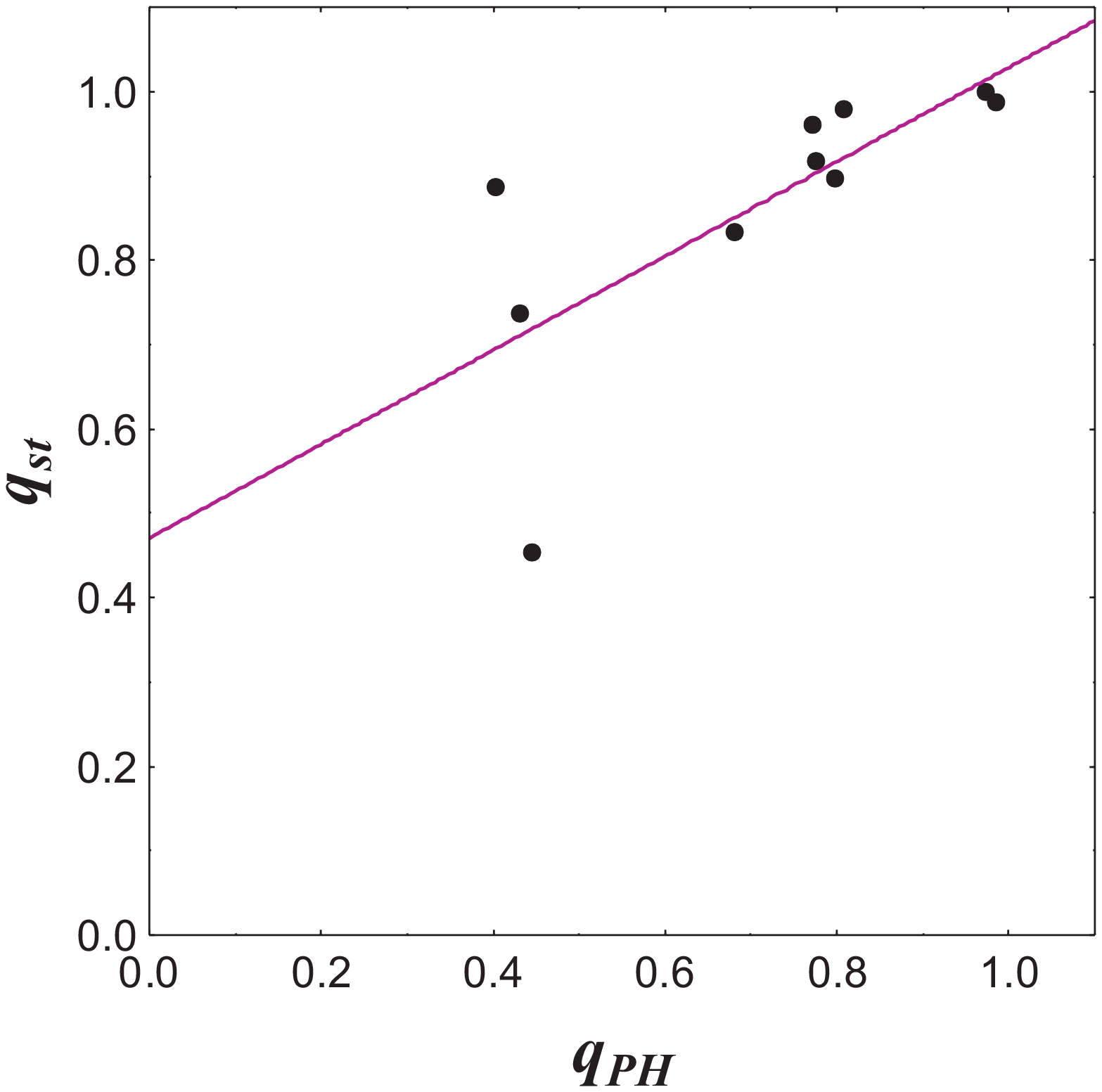}
\caption{Dependence of $r_2/r_1$ on mass ratio \emph{q} (left
        panel), dependence of $T_2/T_1$ on mass ratio \emph{q} (middle panel) and
diagram $q_{st}$ -- $q_{PH}$ (right panel)}\label{Fig14}
\end{center}
\end{figure*}

(10) The ratios $T_2/T_1$ of our targets fall in the range
0.8--1.0 with one exception, KIC 6220470, which value is 0.56
(Fig. 14, middle). Opposite discrepancy has been found for HD 174884
(Maceroni et al. 2009) which components are with equal
temperatures irrespective of the negligible secondary eclipse.

(11) The values $q_{st}$ of the mass ratios, obtained by the
empirical relation temperature--mass of MS stars (Table 4),
differed from those determined by our \emph{PHOEBE} models $q
\equiv q_{PH}$ (Fig. 14, right). The most glaring case is KIC 9474969
with values $q_{st}$=0.88 and $q_{PH}$=0.4.

Figure 14 implies that the components of the eccentric binaries,
especially those with $q \leq 0.5$, do not follow the empirical
relations between the global stellar parameters derived by study
of circular-orbit binaries (and with probable dominance of mass
ratios around unity). In fact only targets KIC 9658118 and KIC
6949550 with mass ratio near unity turned out with almost equal
values of the ratios of luminosities and radii of their components
obtained by the light curve solutions ($l_2/l_1$ and $r_2/r_1$
from Table 2) and by the empirical relations for MS stars
($L_2/L_1$ and $R_2/R_1$ from Table 4). The biggest discrepancies
between the values of these ratios belong to KIC 9474969 with the
smallest mass ratio. This result might be considered as some
empirical support of the conclusion of Sepinsky et al. (2007a)
about the difference of the Roche geometry of circular-orbit and
eccentric-orbit binaries. Another possible reason could be that
the empirical relations for MS stars have been derived on the base
of binaries with mass ratios around unity.

\begin{figure*}
\begin{center}
\includegraphics[width=5cm]{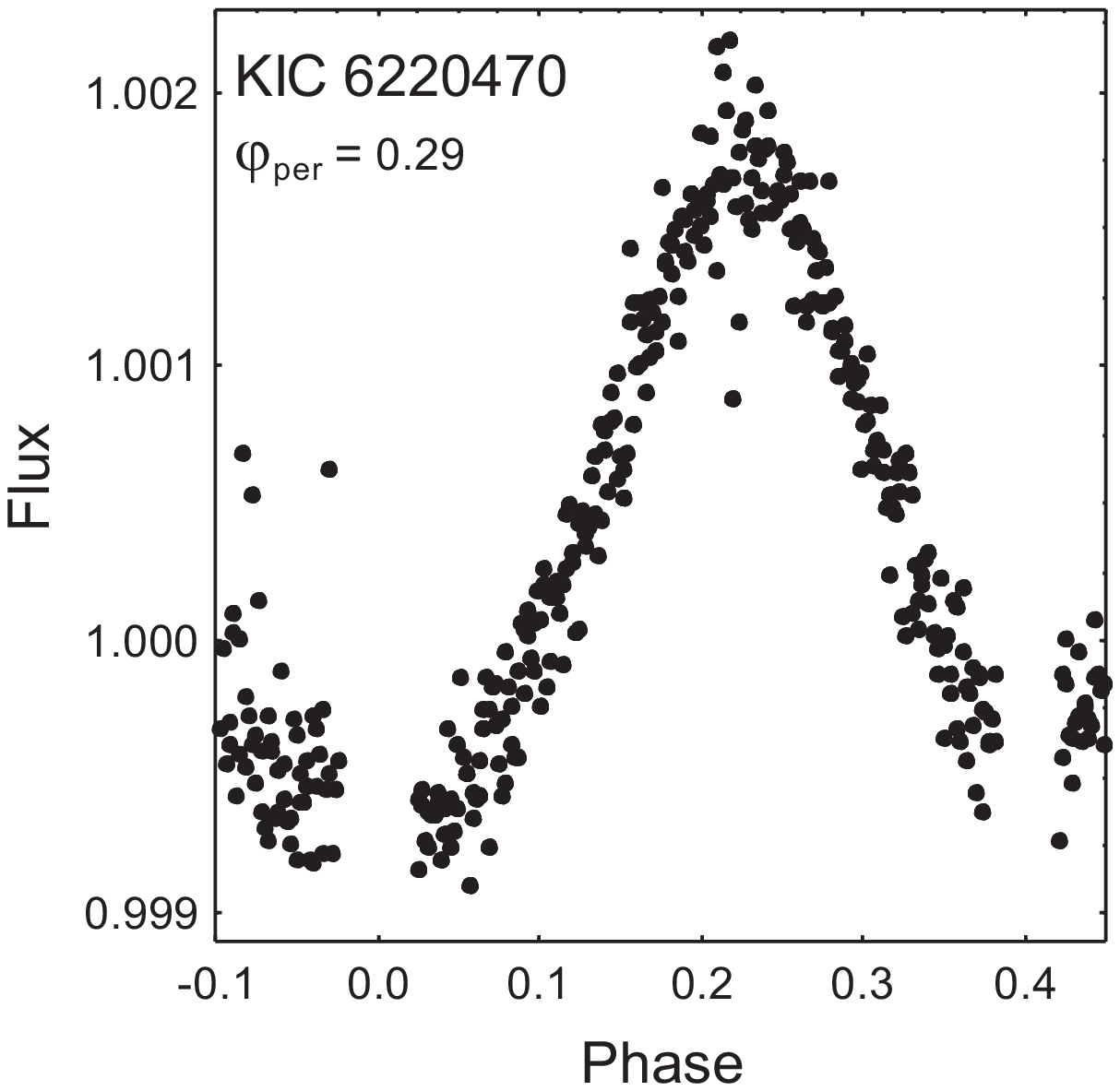}
\includegraphics[width=5cm]{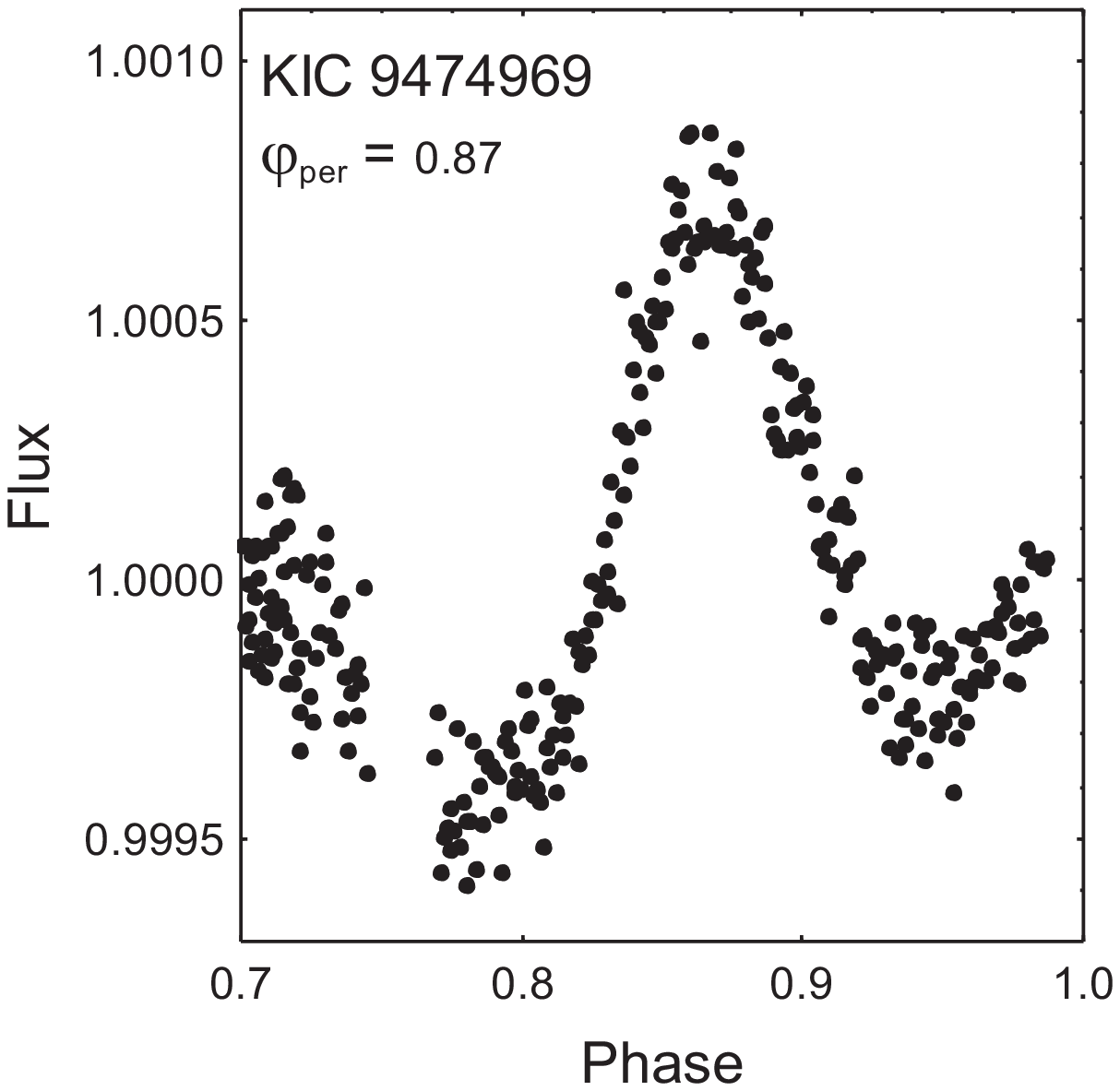}
\includegraphics[width=5cm]{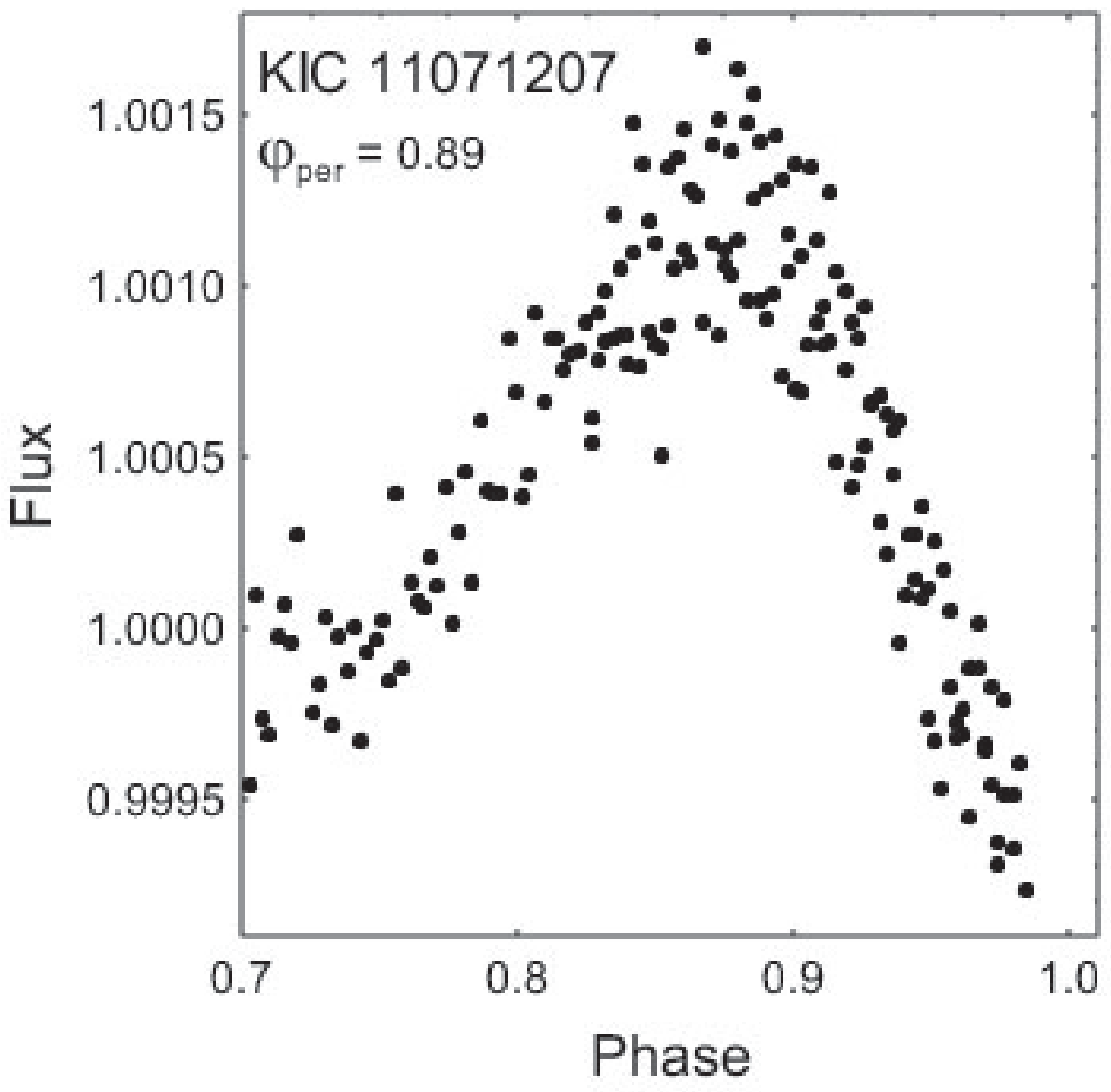}
\caption{Tidally induced brightening of KIC 6220470 (left panel), KIC 9474969 (middle panel) and KIC 11071207 (right panel)}\label{Fig15}
\end{center}
\end{figure*}

\section{TIDALLY INDUCED HUMPS}

The tidal forces change the stellar shape (tidal bulges) and cause
brightness variability due to projection of the distorted stellar
surfaces on the visible plane (Brown et al. 2011; Welsh et al.
2011; Morris 1985). It has double-wave shape (ellipsoidal
variations) in the case of circular orbits and light increasing
around the periastron in the case of eccentric orbits.

Kumar et al. (1995) created an analytic model of tidal phenomena
in eccentric binary consisting of point source (neutron star) and
MS star. The shape of the corresponding light curve depends on
inclination, angle of periastron, and eccentricity while its
amplitude (of order of mmag) depends on the masses of the objects,
their internal stellar structure and the orbital separation at the
peristron (Kumar et al. 1995). According to this model the shape
of the light increase is one-peaked for $i \leq 30^0$ but becomes
two-peaked with central dip (which depth and width increase with
\emph{i}) for the bigger orbital inclination.

Thompson et al. (2012) calculated a grid of solutions to the model
of Kumar et al. (1995) and found that the larger inclinations
cause the light curve firstly to increase in brightness and then
to decrease (KIC 3547874), or vice versa (KIC 9790355), depending
on the angle of periastron, while the bigger eccentricity led to
shorter duration of the heartbeat event.

Three targets from our sample, KIC 6220470, KIC 9474969 and KIC
11071207 (Fig. 15), reveal light features around the periastron
phase $\varphi_{per}$ (Table 2). The analysis of these observed
features led us to the following results.

(a) The observed features around the periastron phase (Fig. 15)
seem as a ''hump'' (brightening). Their shapes differ from the
expected ones for the big orbital inclinations (with central dips,
see figs. 2--3 of Kumar et al. 1995 and fig. 5 of Thompson et al.
2012).

(b) The hump duration is shortest for KIC 9474969 which
eccentricity is above 2 times bigger than those of the other two
''humped'' targets (see Table 2 and Fig. 15). This result supports
the conclusion of Thompson et al. (2012).

(c) The tops of the humps do not coincide perfectly with the
periastron phases: those of KIC 6220470 and KIC 11071207 slightly
precede periastron phase while that of KIC 9474969 slightly delays
(Fig. 15). These small deviations (phase shifts below 0.04)
probably are due to the irradiation effect which phase
contribution depends on the periastron angle.

(d) Humps were found only for those targets from our sample for
which $r_1+r_2 \geq 0.09$ (Table 2). This implied that
the hump amplitude depends strongly on the relative stellar radii.
An additional argument was that the hump amplitude was biggest for
KIC 6220470 (Fig. 15) which value $r_1+r_2 = 0.13$ was the biggest
one (Table 3) among the three ''humped'' targets of our sample.

To derive the dependence of the hump amplitude on the relative
radii we applied the formalism of Kumar et al. (1995) to eccentric
binary consisting of two MS stars and obtained the expression
\begin{equation}
\frac {\delta F }{F}  = 4 \frac {1}{q} \frac {r_2^3}{(1-e)^3} + 4q \frac {r_1^3}{(1-e)^3}  .
\end{equation}
where\emph{ F} is the total flux of the target. In fact, this
formula gives the amplitude of the ''clean'' tidally induced hump.
The irradiation effect, which contribution depends on the
periastron angle, superposes the tidally excited brightening and
changes the hump amplitude. Moreover, possible tidally-induced
pulsations (Willems $\&$ Aerts 2002) and Doppler boosting (Bloemen
et al. 2011) could cause additional complications.

The hump amplitudes of our targets calculated by the expression
(1) were around 2 times bigger than the observed ones. Kumar et
al. (1995) obtained just the same discrepancy for PSR 0045-7319.
This led us to the supposition about wrong coefficient of their
expressions (2 instead of 4).

We should point out that KIC 6949550 has $r_1+r_2$=0.1 but does
not reveal a hump. We attributed this exception from the rule
(existence of detectable hump for $r_1+r_2 \geq 0.09$) by its
semi-regular variations with bigger amplitude which blur the
possible hump.

We assume that the obtained limit of 0.09 of $r_1+r_2$ depends on
the data precision and can be reduced for future space missions.

(e) All targets with periastron brightening have $q \simeq 0.45$.
We attributed this fact as a consequence of the accidental
coincidence of the binaries with $q \simeq 0.45$ and the systems
with biggest relative radii because the dependence (1) of $\delta
F / F$ on the relative radii is stronger than that on the mass ratio.

(f) Humps would not be discern for systems which periastron angle
$\omega$ is close to 90$^0$ or 270$^0$ because the periaston phase
for these cases will coincide with some of the eclipses. Our
targets are not such cases (Table 2) and thus, we have not missed
some humps on this reason.

(g) We did not find oscillatory modes of any target. This result
confirms the conclusion of Kumar et al. (1995) that the MS stars
have small amplitudes of oscillation.

\section{CONCLUSION}

This paper presents the results of determination of the orbits and
fundamental parameters of ten eclipsing binaries with eccentric
orbits from the \emph{Kepler} archive. The obtained results
allowed us to derive several conclusions.

(1) The formal errors of the derived parameters from the light
curve solutions are below 1 $\%$ (excluding those of KIC 6220470
which exceed 3 $\%$).

(2) Our light curve solutions imply that the components of the
eccentric binaries (especially those with mass ratios below 0.5)
do not follow precisely the empirical relations between the stellar
parameters derived from the study of circular-orbit binaries.

(3) KIC 6949550 reveals semi-regular light variations with an
amplitude around 0.004 and a period around 7 d which are modulated
by long-term variations.

(4) We found tidally induced light ''hump'' around the periastron
phase of three targets: KIC 6220470, KIC 11071207 and KIC 9474969.

(5) We derived formula describing the amplitude of the tidally
induced hump of eccentric binary consisting of MS stars. It
exhibits that the amplitude of these features increases strongly
with the relative stellar radii.

(6) Although we did not find evidences of apsidal motion of our
targets in the framework of the relative short duration of the
\emph{Kepler} observations our EEBs present appropriate targets
for future study of this effect due to their big eccentricities.

Obviously, the numerous and exclusive precise \emph{Kepler} data
deserve precise light curve solutions to enrich statistics of the
binaries with estimated parameters and thus to improve the
empirical relations between them. Moreover, the tidally induced
phenomena provide critical data to constrain the theories of tidal
forces in stellar binaries (Fuller $\&$ Lai 2011, Burkart et al.
2012).

%We suppose our results could be incorporated in such studies because our eccentric binaries are with high orbital inclinations
%and hence, their parameters are determined by high precision.

\section*{Acknowledgments}
The research was supported partly by funds of project RD-08-285 of
Scientific Foundation od Shumen University. It used the SIMBAD
database and NASA Astrophysics Data System Abstract Service. We
worked with the live version of the \emph{Kepler} EB catalog.

%from http://keplerEBs.villanova.edu.}

This publication makes use of data products from the Two Micron
All Sky Survey, which is a joint project of the University of
Massachusetts and the Infrared Processing and Analysis
Center/California Institute of Technology, funded by the National
Aeronautics and Space Administration and the National Science
Foundation (Skrutskie et al. 2006). This research makes use of the
SIMBAD and Vizier data bases, operated at CDS, Strasbourg, France,
and NASA Astrophysics Data System Abstract Service.

The authors are grateful to anonymous referee for the valuable
notes and propositions.


\begin{thebibliography}{}

\bibitem[\protect\citename{Barembaum }1995]{bar}
Barembaum Morrie J., Etzel Paul B., 1995, AJ, 109, 2680

\bibitem[\protect\citename{Bate }1997]{bat1}
Bate M., 1997, MNRAS, 285, 16

\bibitem[\protect\citename{Bate }1997]{bat2}
Bate M., Bonnell I., 1997, MNRAS, 285, 33

\bibitem[\protect\citename{Bate }2002]{bat3}
Bate M., Bonnell I., Bromm  V., 2002, MNRAS, 336, 705

\bibitem[\protect\citename{Bloemen et al. }2011]{blo}
Bloemen, S., et al. 2011, MNRAS, 410, 1787

\bibitem[\protect\citename{Bonnell }1992]{bon}
Bonnell I., Bastien P., 1992, ApJ, 401, 654

\bibitem[\protect\citename{Borkovits }2014]{bor}
Borkovits T. et al., 2014, MNRAS, 443, 3068

\bibitem[\protect\citename{Boyajian et al. }2013]{boy}
Boyajian T. et al., 2013, ApJ, 771,

\bibitem[\protect\citename{Brown et al. }2011]{bro}
Brown, W. R., Kilic, M., Hermes, J. J., Allende Prieto, C.,
Kenyon, S. J., Winget, D. E. 2011, ApJ, 737, L23

\bibitem[\protect\citename{Bulut }2007]{bul1}
Bulut, I.; Demircan, O., 2007, MNRAS,378, 179

\bibitem[\protect\citename{Bulut et al. }2014]{bul2}
Bulut, I., Bulut, A., Cicek, C., 2014, NewA, 32, 21

\bibitem[\protect\citename{Burkart et al. }2012]{bur}
Burkart, J., Quataert, E., Arras, P., Weinberg, N. N., 2012,
MNRAS, 421, 983

\bibitem[\protect\citename{Claret }1991]{cla1}
Claret A., Gimenez A., 1991, A \& A, 244, 319

\bibitem[\protect\citename{Claret }1993]{cla2}
Claret A., Gimenez A., 1993, A \& A, 277, 487

\bibitem[\protect\citename{Claret }2010]{cla3}
Claret  A., Gimenez A., 2010, A \& A, 519A, 57

\bibitem[\protect\citename{Claret }2011]{cla4}
Claret A., Bloemen S., 2011, A \& A,529A, 75

\bibitem[\protect\citename{Claret }2012]{cla5}
Claret A., 2012, A \& A, 541A, 113

\bibitem[\protect\citename{Coughlin et al. }2011]{cou}
Coughlin, J. L.; Lopez-Morales, M.; Harrison, T. E.; Ule, N.; Hoffman, D. I., 2011, AJ, 141, 78

\bibitem[\protect\citename{De Cat et al. }2000]{dec}
De Cat, P., Aerts, C., De Ridder, J., et al. 2000, A \& A, 355,
1015

\bibitem[\protect\citename{Dimitrov et al. }2012]{dim1}
Dimitrov D., Kjurkchieva D., Radeva V., 2012, BlgAJ 18c, 81

\bibitem[\protect\citename{Dimitrov }2015]{dim2}
Dimitrov D., Kjurkchieva D., 2015, MNRAS, 448, 2890

\bibitem[\protect\citename{Duquennoy }1991]{duq}
Duquennoy A., Mayor M., 1991, A \& A, 248, 485

\bibitem[\protect\citename{Fuller }2012]{ful}
Fuller, J., Lai, D., 2012, MNRAS, 420, 3126

\bibitem[\protect\citename{Garcia et al. }2014]{gar}
Garcia, E. V.; Stassun, Keivan G.; Pavlovski, K.; Hensberge, H.;
Gómez Maqueo Chew, Y.; Claret, A., 2014, AJ, 148, 39

\bibitem[\protect\citename{Gimenez }1985]{gim1}
Gimenez A., 1985, ApJ, 297, 405

\bibitem[\protect\citename{Gimenez }1992]{gim2}
Gimenez  A., Quintana  J. M., 1992, A \& A, 260, 227

\bibitem[\protect\citename{Graczyk }2003]{gra}
Graczyk D., 2003, MNRAS, 342, 1334

\bibitem[\protect\citename{Gundlach }2011]{gun}
Gundlach Carsten, Murphy Jeremiah W., 2011, MNRAS, 416, 1284

\bibitem[\protect\citename{Hambleton et al. }2013]{ham}
Hambleton, K. M.; Kurtz, D. W.; Prsa, A.; Guzik, J. A.; Pavlovski,
K.; Bloemen, S.; Southworth, J.; Conroy, K.; Littlefair, S. P.;
Fuller, J., 2013, MNRAS, 434, 925

\bibitem[\protect\citename{Handler et al. }2002]{han}
Handler, G., et al. 2002, MNRAS, 333, 262

\bibitem[\protect\citename{Harmanec et al. }2014]{har}
Harmanec, P. et al., 2014, A \& A, 563A, 120

\bibitem[\protect\citename{Hernandez }2011]{her}
Hernandez-Gomez, A. et al., 2011, RMxAC, 40, 278

\bibitem[\protect\citename{Kipping }2010]{kip}
Kipping, D. M., 2010, MNRAS, 408, 1758

\bibitem[\protect\citename{Kjurkchieva }2015]{kju}
Kjurkchieva D., Dimitrov D., 2015, AN, accepted

\bibitem[\protect\citename{Koch et al. }2010]{koch}
Koch D. G., et al. 2010, ApJ, 713, L79

%\bibitem[\protect\citename{Koenigsberger }2009]{koe} Koenigsberger Gloria, Moreno Edmundo, 2009, arXiv, 0903, 1221

\bibitem[\protect\citename{Kopal }1978]{kop}
Kopal Z., 1978, ASSL (Astrophysics and Space Science Library) 68,
Dordrecht, D. Reidel Publishing Co.

\bibitem[\protect\citename{Kozyreva }2014]{koz}
Kozyreva V. S., Kusakin  A. V., 2014, Ap, 57, 221

\bibitem[\protect\citename{Kumar et al. }1995]{kum}
Kumar P., Ao C., Quataert E., 1995, ApJ 449, 294

\bibitem[\protect\citename{Kuznetsov et al. }2011]{kuz}
Kuznetsov M.V. et al., 2011, ARep, 55, 989

\bibitem[\protect\citename{Lacy et al. }2015]{lac}
Lacy Claud H. Sandberg; Torres Guillermo; Fekel Francis C.; Muterspaugh Matthew W.; Southworth, John, 2015, AJ, 149,34

\bibitem[\protect\citename{Lajoie }2011]{laj}
Lajoie Charles-Philippe, Sills Alison, 2011, ApJ, 726, 67

\bibitem[\protect\citename{Lehmann et al. }2013]{leh}
Lehmann, H.; Southworth, J.; Tkachenko, A.; Pavlovski, K., 2013, A
\& A, 557A, 79

\bibitem[\protect\citename{Levi-Civit }1937]{lev}
Levi-Civita T., 1937, Amer. J. Math., 59, 225

\bibitem[\protect\citename{Lucy }2006]{luc}
Lucy L., 2006, A \& A, 457, 629

\bibitem[\protect\citename{Maceroni et al. }2009]{mac1}
Maceroni, C., et al. 2009, A \& A, 508, 1375

\bibitem[\protect\citename{Maceroni et al. }2014]{mac2}
Maceroni, C.; Lehmann, H.; da Silva, R.; Montalba'n, J.; Lee,
C.-U.; et al., 2014, A \& A, 563A, 59

\bibitem[\protect\citename{Mathieu }1988]{mat}
Mathieu, R. D., Mazeh, T., 1988, ApJ, 326, 256

\bibitem[\protect\citename{Michalska }2004]{mich1}
Michalska G., Pigulski A., 2004, New Astronomy Rev., 48, 719

\bibitem[\protect\citename{Michalska }2005]{mich2}
Michalska G., Pigulski A., 2005, A \& A, 434, 89

\bibitem[\protect\citename{Michalska }2007]{mich3}
Michalska G., 2007, IBVS, 5759

\bibitem[\protect\citename{Morris }1985]{mor}
Morris, S. L. 1985, ApJ, 295, 143

\bibitem[\protect\citename{North et al. }2010]{nor}
North, P.; Gauderon, R.; Barblan, F.; Royer, F., 2010, A \& A,
520A, 74

\bibitem[\protect\citename{Petrova }1999]{petr}
Petrova, A.V., Orlov, V.V. 1999, AJ, 117, 587

\bibitem[\protect\citename{Pichardo et al. }2005]{pic}
Pichardo B., Sparke L., Aguilar L., 2005, MNRAS, 359, 521

\bibitem[\protect\citename{Prsa }2005]{prs1}
Prsa A., Zwitter T., 2005, ApJ 628, 426

\bibitem[\protect\citename{Prsa et al. }2008]{prs2}
Prsa A. et al., 2008, ApJ 687, 542

\bibitem[\protect\citename{Prsa et al. }2011]{prs3}
Prsa A. et al., 2011, AJ 141, 83

\bibitem[\protect\citename{Sepinsky et al. }2007]{sep1}
Sepinsky J.F., Willems B., Kalogera V. 2007a, ApJ, 660, 1624

\bibitem[\protect\citename{Sepinsky et al. }2007]{sep2}
Sepinsky J.F., Willems B., Kalogera V., Rasio F.A., 2007b, ApJ,
667, 1170

\bibitem[\protect\citename{Sepinsky et al. }1990]{sep3}
Sepinsky J.F., Willems B., Kalogera V., Rasio F.A., 2009, ApJ,
702, 1387

\bibitem[\protect\citename{Slawson et al. }2011]{sla}
Slawson R. et al., 2011, AJ 142, 160

\bibitem[\protect\citename{Skrutskie et al.}{2006}]{skr}
Skrutskie M. F., Cutri R. M., Stiening R., Weinberg M. D.,
Schneider S., Carpenter J. M., Beichman C., Capps R. et al. 2006,
AJ, 131, 1163

\bibitem[\protect\citename{Song et al. }{2013}]{song}
Song H. F. et al., 2013, A \& A, 556A, 100

\bibitem[\protect\citename{Terrell }2005]{ter}
Terrell D., Wilson R., 2005, Ap \& SS, 296, 221T

\bibitem[\protect\citename{Thompson et al. }2012]{tho}
Thompson S. E. et al., 2012, ApJ, 753, 86

\bibitem[\protect\citename{Welsh W. et al. }2011]{wel}
Welsh W. et al., 2011, ApJS, 197, 4

\bibitem[\protect\citename{Willems }2002]{wil1}
Willems, B., Aerts, C. 2002, A \& A, 384, 441

\bibitem[\protect\citename{Willems }2003]{wil2}
Willems, B. 2003, MNRAS, 346, 968

\bibitem[\protect\citename{Willems }2005]{wil3}
Willems B., Claret A., 2005, SPC, 333, 52W

\bibitem[\protect\citename{Willems }2007]{wil4}
Willems B., 2007, ASPC, 361, 124

\bibitem[\protect\citename{Wilson }1971]{wils1}
Wilson R. E., Devinney E. J., 1971, ApJ, 166, 605

\bibitem[\protect\citename{Wilson }1979]{wils2}
Wilson R. E., 1979, ApJ, 234, 1054

\bibitem[\protect\citename{Wilson }2004]{wils3}
Wilson R. E., Van Hamme W., 2004, Computing Binary Star Observables (in Reference Manual to the Wilson-
Devinney Program)

\bibitem[\protect\citename{Wilson }2014]{wils4}
Wilson, R. E., Van Hamme, W., 2014, ApJ, 780, 151

\bibitem[\protect\citename{Wolf et al. }2010]{wol1}
Wolf M., Claret A., Kotkova L., Kucakova H., Kocian R., Brat L., Svoboda P., Smelcer L., 2010, A \& A, 509A, 18

\bibitem[\protect\citename{Wolf et al. }2013]{wol2}
Wolf, M., Zasche, P., Kucakova, H., Lehky, M., Svoboda, P.,
Smelcer, L., Zejda, M., 2013, A \& A, 549A,108

\bibitem[\protect\citename{Zahn }2005]{zahn}
Zahn, J.-P. 2005, in Tidal Evolution and Oscillations in Binary
Stars, ed. A. Claret, A. Giménez, J.-P. Zahn, ASP Conf. Ser., 333,
4

\bibitem[\protect\citename{Zasche }2012]{zas1}
Zasche  P., 2012, AcA, 62, 97Z

\bibitem[\protect\citename{Zasche }2013]{zas2}
Zasche P., Wolf M., 2013, A \& A 559A, 41

\bibitem[\protect\citename{Zasche et al. }2014]{zas3}
Zasche P., Wolf M., Vrastil J., Liska J., Skarka M., Zejda M.,
2014, A \& A  572A, 71

\end{thebibliography}
\end{document}